\documentclass[twocolumn,preprintnumbers,amsmath,amssymb,aps,pra,longbibliography]{revtex4-1}
\usepackage[T1]{fontenc}
\usepackage{amssymb,amsmath,bm}
\usepackage{graphicx}
\usepackage{color}
\usepackage{bbold}
\usepackage{bbm}
\usepackage{appendix}
\usepackage{soul}
\usepackage[dvipsnames]{xcolor}
\usepackage{ulem}
\usepackage{float}
\usepackage{textcomp}
\usepackage{ mathrsfs }
\usepackage{siunitx}%
\usepackage{braket}

\begin{document}
\title{Exciton-Polariton hybrid skin-topological states}
\author{Ruiqi Bao$^{1,2}$}
\email[Corresponding author:~]{ruiqi002@e.ntu.edu.sg}
\author{R. Banerjee$^{1}$}
\author{S. Mandal$^{3}$}
\author{Huawen Xu$^{4}$}
\author{Shiji Li$^{2}$}
\author{Junfeng Gao$^{2}$}
\email[Corresponding author:~]{gaojf@dlut.edu.cn}

\author{Timothy C. H. Liew$^{1}$}\email[Corresponding author:~]{timothyliew@ntu.edu.sg}
\affiliation{$^1$Division of Physics and Applied Physics, School of Physical and Mathematical Sciences, Nanyang Technological University, Singapore 637371, Singapore \\$^2$Laboratory of Materials Modification by Laser, Ion and Electron Beams, Ministry of Education, Dalian University of Technology, Dalian, 116024 China\\ $^3$ Department of Physics, Indian Institute of Technology Bombay, Mumbai 400076, India \\ $^4$ Beijing Academy of Quantum Information Sciences, Beijing 100193,
China  }


\begin{abstract}

The non-Hermitian skin effect (NHSE), where bulk states accumulate at system boundaries, challenges the conventional bulk-boundary correspondence. Here we propose a scheme to realize hybrid skin–topological states (HSTS) in exciton polariton honeycomb lattices by introducing sublattice-dependent gain and loss. This non-Hermiticity couples with the intrinsic topological edge modes, leading to relocalization of edge states. We show two distinct regimes: hybrid skin–Chern states with switchable localization controlled by TE–TM splitting (characterized by a change in spectral winding number from $w=2$ to $w=-1$), and hybrid skin–antichiral states which preserves the spin-polarized property. Our results bridge polariton spin physics and non-Hermitian topology, opening routes toward controllable non-reciprocal and spin-polarized transport.


\end{abstract}

\maketitle
\emph{\textcolor{red}{Introduction.}}
Non-Hermitian Hamiltonians are used to describe open quantum systems especially those having dissipation and gain. Non-Hermitian systems have attracted intense research interest in recent years, particularly due to their ability to host new topological phases that have no counterparts in Hermitian systems~\cite{Feng:2017aa,PhysRevLett.118.040401,El-Ganainy:2018aa,Ding:2022aa}. 
Among them, the non-Hermitian skin effect (NHSE), where bulk states accumulate at the edge of a finite system,
violates the conventional bulk-boundary correspondence (BBC) in Hermitian topological cases~\cite{PhysRevLett.121.086803,PhysRevLett.125.186802,PhysRevLett.124.086801,PhysRevLett.123.246801,PhysRevLett.121.026808,PhysRevLett.123.066404,PhysRevLett.125.226402,PhysRevLett.125.126402} and stimulates research across photonics~\cite{doi:10.1126/science.aaz8727}, acoustics, and electronic platforms~\cite{doi:10.1073/pnas.2010580117,Zhang:2021aa,doi:10.34133/2021/5608038,PhysRevLett.129.070401,Li:2024aa,LIU20241228,Li:2020aa,Zhang:2021ab,PhysRevLett.123.170401,PhysRevLett.124.250402}. 
 
Building upon these advancements, recent research has turned to explore the effects of the NHSE on topologically protected states, aiming to understand the interplay between topology and non-Hermiticity.
The interplay between the NHSE and one-dimensional topological states or defect states has shown its ability to manipulate the spatial distribution of states which is called non-Hermitian morphing~\cite{Wang:2022vc,PhysRevB.108.235305}. 
In Chern insulators, the coexistence of chiral edge states and the NHSE gives rise to hybrid skin-topological states (HSTS), where chiral edge modes become spatially localized even though the Chern number remains non-trivial ~\cite{PhysRevLett.123.016805,PhysRevB.102.205118,PhysRevB.106.035425,PhysRevLett.128.223903,Zhu_2024,PhysRevB.108.075122}. Despite these advances, HSTS have thus far been explored primarily in low-frequency systems such as acoustics, electrical circuits, and microwaves~\cite{fang2025tunablehingeskinstates,Zou:2021aa,Jiang:2024aa,PhysRevLett.132.113802}. It is essential to investigate such phenomena at optical frequencies, where strong light–matter interactions can enable functional device architectures at the nano- to microscale. Furthermore, the influence of non-Hermiticity on antichiral edge states and on internal degrees of freedom such as optical spin remains largely unexplored, offering additional avenues for realizing novel non-Hermitian optical devices.

Exciton-polaritons, hybrid light-matter quasiparticles formed in the strong coupling regime between cavity photons and excitons~\cite{RevModPhys.85.299,RevModPhys.82.1489,Kavokin:2022aa}, provide an ideal platform to explore these effects. 
Because of the exciton part, polaritons can react to external magnetic field, which leads to Zeeman splitting and breaks the time reversal (TR) symmetry. Further combined with transverse electric-transverse magnetic (TE-TM) splitting from the photon part, polaritons are known to host Chern insulators and antichiral edge states~\cite{PhysRevB.91.161413,PhysRevLett.114.116401,Klembt:2018wu,PhysRevB.99.115423,PhysRevB.106.235310}.
Polaritons are driven-dissipative systems, which means that their gain can be controlled by external non-resonant pumps. The gain/loss control makes polaritons a great platform to study non-Hermitian physics such as exceptional points~\cite{PhysRevLett.120.065301,Opala:23,PhysRevB.109.085311,PhysRevResearch.6.013148,Li:2022aa}, parity-time (PT) symmetry~\cite{Ma_2019,PhysRevB.109.085312} and also the NHSE~\cite{PhysRevLett.125.123902,PhysRevB.104.195301,Mandal:2022aa,PhysRevB.108.235305,PhysRevB.108.L041403}. Although several theoretical proposals have been proposed to realize the NHSE in polariton system,
they mostly rely on spin-dependent losses that are experimentally challenging to maintain due to spin relaxation in the exciton reservoir~\cite{Carlon-Zambon:2019aa}.

In this work, instead of relying on spin, we theoretically propose a way to realize the NHSE by considering sublattice dependent decay rates. We demonstrate two distinct types of HSTS: hybrid skin-Chern edge states (HSCS) and hybrid skin-antichiral edge states (HSAS). First, we demonstrate HSCS are relocalized due to the NHSE and exhibit switchable localization characterized by spectral winding numbers. By tuning the strength of the TE-TM splitting, we drive a topological phase transition in the underlying Hermitian system (from Chern number $C= 2$ to $C=1$), and reverse the localization direction of the HSCS. The switchable localization offers a novel mechanism for reconfiguring topological states. Second, we reveal the spin-polarized HSAS. These states are localized towards  opposite directions due to the NHSE, and they inherit their spin-polarized property from their parent antichiral edge states. This results in a robust spatial separation of spin-up and spin-down channels, which is confirmed by the non-trivial spectral winding numbers. Our work provides a way to create and control distinct HSTS, which is vital for non-reciprocal photonic transport and topological lasing.

\emph{\textcolor{red}{Model.}} \label{sec:model}
\begin{figure}
\includegraphics[width=1\columnwidth]{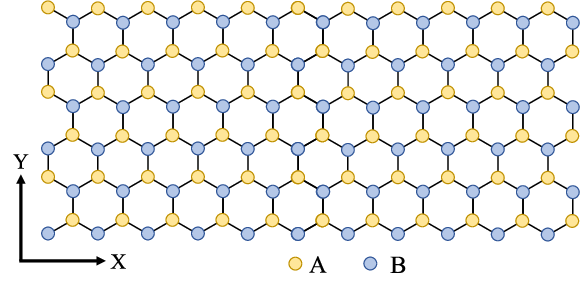}
\caption{ Schematic figure of the non-Hermitian honeycomb lattice formed by two types of sublattices shown in yellow and blue colors separately. To make the system non-Hermitian, we set sublattice-dependent gain/loss. }
\label{Fig1}
\end{figure}
We start by considering a honeycomb lattice formed by two sublattices A and B in the presence of Zeeman splitting and TE-TM splitting as shown in Fig. 1.
In this work, we investigate how the non-Hermitian gain/loss will influence the behaviour of the two kinds of topological states.

Polaritons correspond to a natural driven-dissipative system and in principle the loss should be the same on different sublattices (say $-i\gamma_0$).
However the loss can be compensated by applying a non-resonant (effectively incoherent) pump, which serves as gain (say $iP_{A,B}$).
By applying different strengths of incoherent pumps on different sublattices, the system becomes non-Hermitian ($-i\gamma_{A,B} = iP_{A,B}-i\gamma_0$).
Taking periodic boundary conditions on both directions, the corresponding Hamiltonian in reciprocal space can be written in the basis of $\Psi = [\Psi_A^+,\Psi_A^-,\Psi_B^+,\Psi_B^-]^T$ as:

\begin{align}\label{Eq2}
\mathcal{H}_{\mathbf{k}}
=\begin{bmatrix}
            \Delta_A-i\gamma_A & 0  & -g_{\mathbf{k}}J  & -g^+_{\mathbf{k}}\Delta_T \\
            0   & -\Delta_A-i\gamma_A &-g^-_{\mathbf{k}}\Delta_T & -g_{\mathbf{k}}J  \\
            -g^*_{\mathbf{k}}J &-{g^-_\mathbf{k}}^*\Delta_T& \Delta_B-i\gamma_B & 0 \\
            -{g^+_\mathbf{k}}^*\Delta_T & -g^*_{\mathbf{k}}J & 0 & -\Delta_B-i\gamma_B
\end{bmatrix},
\end{align}  
where $g_{\mathbf{k}}=\sum_{n=1}^3\exp\left({-i\mathbf{k}.{\mathbf{r}}_n}\right)$ and 
$g^\pm_{\mathbf{k}}=\sum_{n=1}^3\exp\left({-i\left[\mathbf{k}.{\mathbf{r}}_n\mp 2\theta_n\right]}\right).$
Here $\mathbf{r}_{n}$ represent the vectors connecting the three nearest "B" sites from a single "A" site (Fig.~\ref{Fig1}) and $\theta_n=2\pi(n-1)/3$ are the angles of those vectors  with respect to one of them (say $\mathbf{r}_{1}$).
In the Hamiltonian, $J$ is the coupling between the nearest neighbouring pillars, $\Delta_T$ is the TE-TM splitting, $\Delta_{A(B)}$ is the onsite energy introduced by Zeeman splitting on respective sites, $\gamma_{A(B)}$ is the effective loss on $A(B)$ sites. Here if $\Delta_{A} = \Delta_{B}$, the system supports polariton Chern insulators, while when $\Delta_{A} = -\Delta_{B}$ the system gives polariton antichiral edge states~\cite{PhysRevB.106.235310}.
\begin{figure}
\includegraphics[width=1\columnwidth]{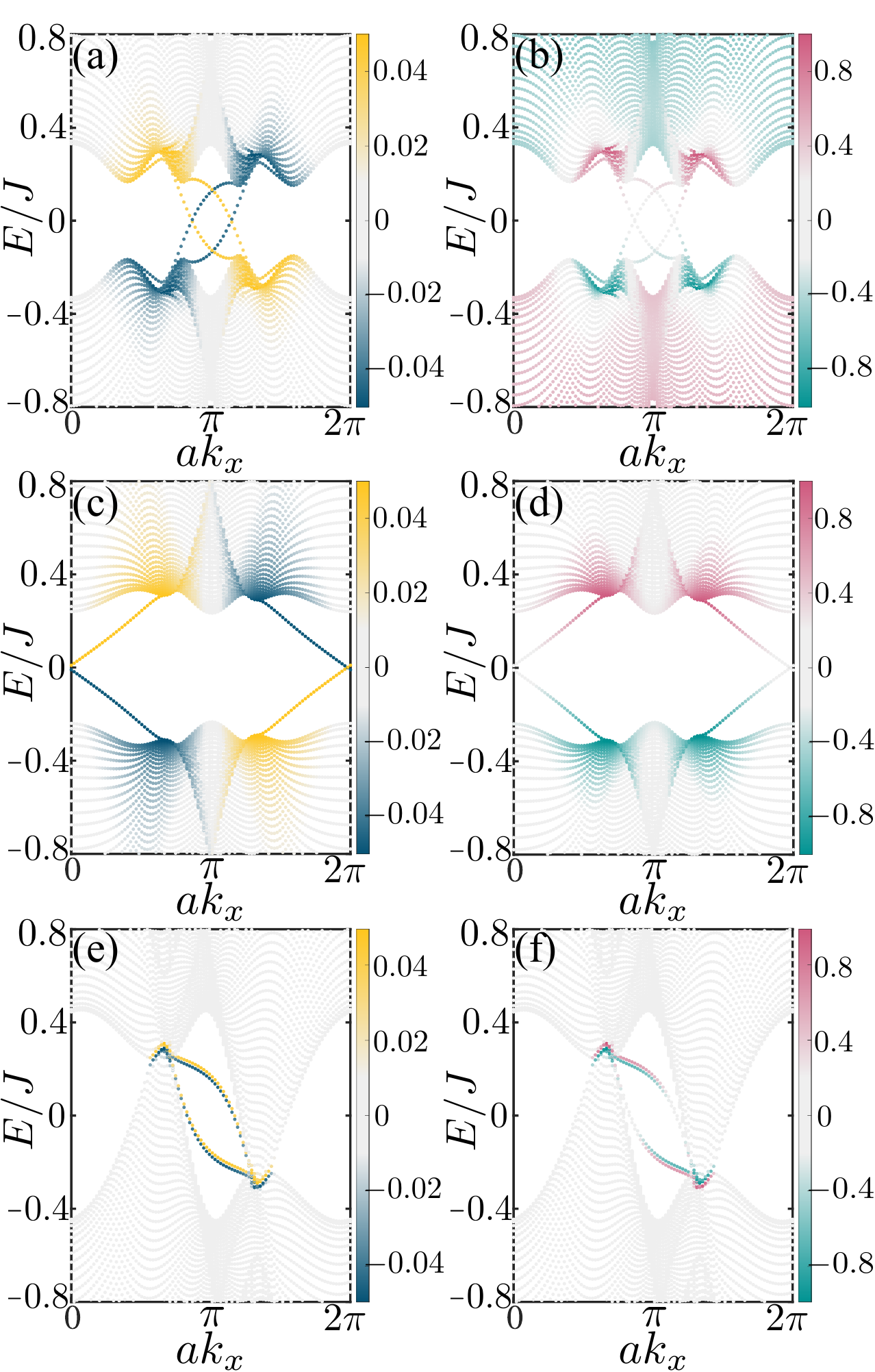}
\caption{ Band structures for Chern insulator $C=2$ (a-b), $C=1$ (c-d) and antichiral edge states (e-f) in non-Hermitian regime. In (a, c, e), the states are color coded according to the imaginary term (gain/loss) and in (b, d, f) are color coded according to the DOCP. In the Chern insulator, the left propagating states and the right propagating states have different signs of imaginary energies and the states are not spin-polarized. For the antichiral case, the edge states are propagating along the same direction and still remain with the spin-polarized properties in the presence of non-Hermiticity.
Parameters: $J = 1, z = 0.3J, \gamma_A = -0.05J, \gamma_B = -\gamma_A$. $\delta J = 0.3J$ for (a, b, e, f) and $\delta J = 0.6J$ for (c, d).}
\label{Fig2}
\end{figure}

Next, we calculate the band structures for the polariton Chern insulator and the antichiral edge states in non-Hermitian regime by taking $x$-periodic boundary condition (PBC)/$y$-open boundary condition (OBC).
To better visualize the properties of topological states, here we color code the band structures according to the imaginary energies (Fig. \ref{Fig2}(a,c,e)) and degree of circular polarization (DOCP)(Fig. \ref{Fig2}(b,d,f)), which is defined as:
\begin{equation}
S_z = \frac{\sum |\Psi_{A,B}^+|^2 - \sum |\Psi_{A,B}^-|^2}{\sum |\Psi_{A,B}^+|^2 + \sum |\Psi_{A,B}^-|^2}.
\end{equation}

In Fig.~\ref{Fig2}(a,b), the band structures for non-Hermitian Chern insulator edge states are shown ($C=2$ case). We find that there are two pairs of edge states existed in the bandgap, consistent with the Chern number. For states with different directions of propagation, they also have different imaginary energies. The DOCP of non-Hermitian topological Chern insulators shows that the edge states are mixed spins and also bulk states are not linearly polarized.

Fig.~\ref{Fig2}(c,d) show the band structures for the case $C=1$ obtained by changing the TE-TM splitting to $\Delta_T = 0.6J$. Only one pair of edge states remains in the band gap, again with opposite imaginary energy having opposite propagation directions. The DOCP distribution in Fig.~\ref{Fig2}(d) confirms that these states still exhibit mixed-spin character.

For the antichiral case, the edge states co-exist in the same energy range as the bulk modes and have the same direction of propagation~\cite{SM1}. We notice that states from different edges have different imaginary parts. The upper edge is positive and the lower edge is negative. More interestingly, we find that the edge states remain circularly polarized and states from the two opposite edges have opposite circular polarization. 

We also calculate the band structures in a continuous model. The results show the change in the number of edge states in the Chern case and spin-polarized property in the antichiral case, which is consistent with our tight binding results (see Supplementary Material~\cite{SM}).

\begin{figure}
\includegraphics[width=1\columnwidth]{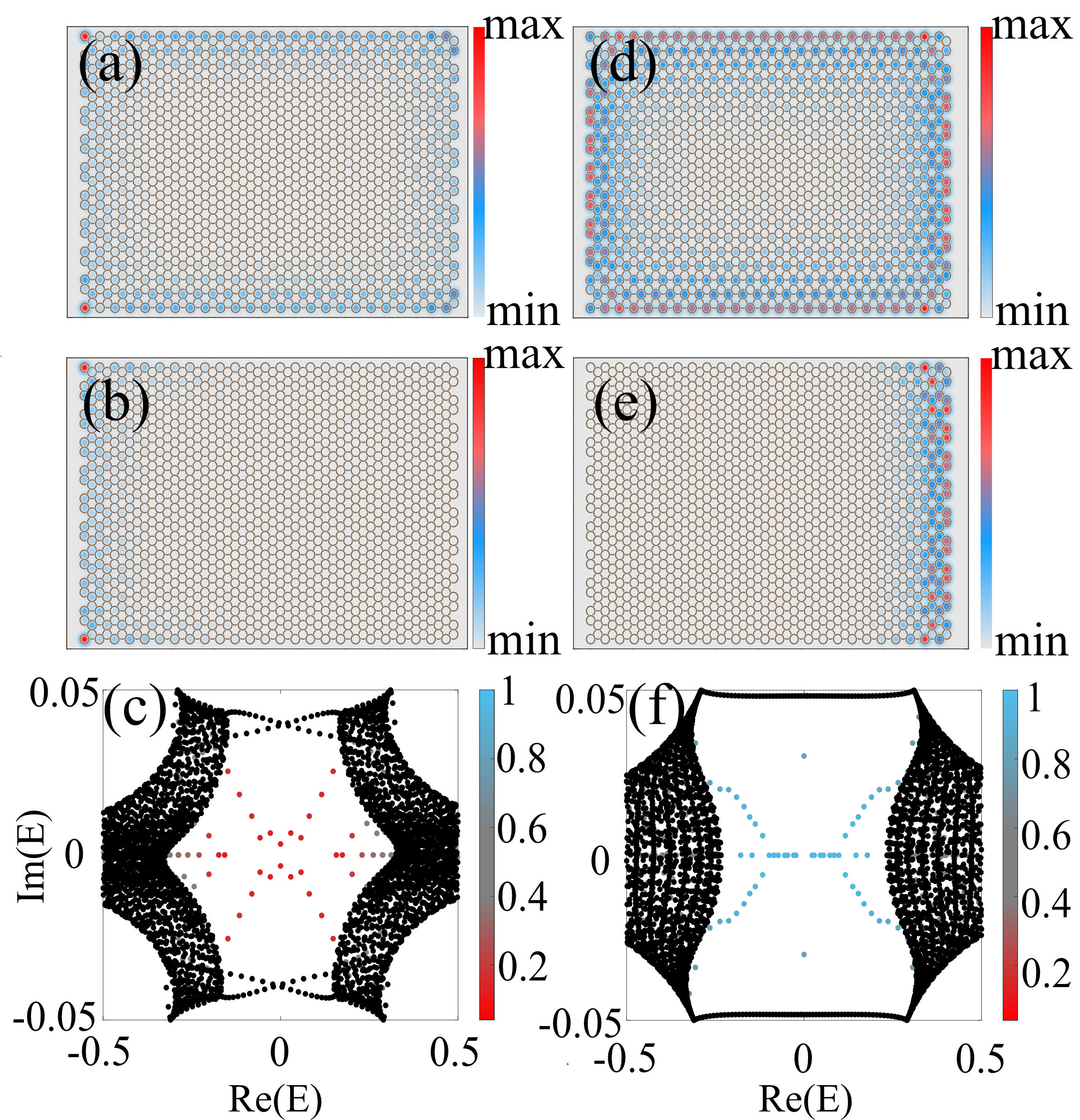}
\caption{(a,d) Spatial distribution of Hermitian Chern edge states. (b,e) Spatial distribution of the HSCS in two cases. The localization direction can be switched  (c,f) Energy spectrum of considered HSCS system for Chern edge states ($C=2,C=1$) separately under both OBCs (color coded according to the expected position) and $x$-PBC/$y$-OBC (black points). Parameters are set the same as in Fig.~\ref{Fig2}.}
\label{Fig3}
\end{figure}

\emph{\textcolor{red}{Switchable hybrid skin-Chern edge states.}}
With insights into the impact of the introduced non-Hermiticity on the band structures, we now explore its influence on the spatial localization of topological edge states. 
We start from the $C=2$ case, by considering OBCs on both directions and calculate the spatial distributions of Hermitian Chern edge states. As shown in Fig.~\ref{Fig3}(a), the edge states are localized all around the boundaries of considered system in the Hermitian case. Upon applying the non-Hermitian condition, the edge states are relocalized towards left boundary (Fig.~\ref{Fig3}(b)). The redistribution of edge states suggests the appearance of hybrid skin-Chern edge states. 

To verify, we calculate the energy spectrum from $x$-PBC/$y$-OBC (black points) and both OBCs. 
We color coded the spectrum by the intensity weighted position (equivalent to the expectation value), which is defined as:
\begin{equation}
\langle x \rangle = (\sum_{x=1}^{N_x}x \cdot |\psi_n(x)|^2)/N_x,
\end{equation}
$N_x$ is the number of sites along the $x$ direction. The expectation values of position approach $1/N_x$ when eigenstates are localized at the left end; and approach 1 when they are localized at the right end.
We find that the energy spectrum changes drastically when the boundary conditions are changed (Fig.~\ref{Fig3}(c)) and the energies of topological edge states from both OBCs are surrounded by the energies from $x$-PBC/$y$-OBC, which suggests the emergence of the NHSE on the edge states. The interplay between NHSE and the Chern edge states leads to the relocalization of topological edge states and forms the HSCS. 
It should be noted that at the same time the bulk states are not localized. The number of HSCS does not scale as $\mathcal{O}$ $(N_xN_y)$; instead it scales as $\mathcal{O}$ $(N_x)$, consistent with previous works~\cite{PhysRevLett.123.016805,PhysRevB.102.205118,PhysRevB.106.035425,PhysRevLett.128.223903,Zhu_2024,PhysRevB.108.075122}. 

The direction of this NHSE-induced localization is explained by the topological invariant (winding number):
\begin{equation}
w = \frac{1}{2\pi i}\int_{BZ} log(\frac{d}{dk}[H(k)-E_{ref}]dk).
\end{equation}
In this case, our calculation yields a winding number of $w=2$. This non-zero integer value provides direct proof of the NHSE and is consistent with the  localization of the HSCS to the left boundary.

Remarkably, the localization of these HSCS is switchable. 
In the Hermitian case, the system undergoes a topological phase transition by tuning the TE-TM splitting ($\Delta_T$). The Chern number changes from $2$ to $1$ when $\Delta_T>J/2$. The edge states of the Hermitian case are still localized all around the boundaries as shown in Fig.~\ref{Fig3}(b). 
Introducing the same non-Hermitian terms now shows a completely different result. The HSCS become localized at the right boundary of the system, as shown in Fig. 3(d).
The reversal of localization can be understood by calculating the winding number which is $-1$ in this case. The change in the sign of the winding number directly corresponds to the direction switching of the HSCS localization from the left to the right boundary. 
This demonstrates a powerful method for reconfiguring topological states, where a tunable parameter can switch not only the underlying Hermitian topology but also localization behaviour of the HSCS. The switchable localization of HSCS is also demonstrated in the continuous model (see Supplementary Material~\cite{SM}).

\emph{\textcolor{red}{Spin-polarized hybrid skin-antichiral edge states.}}
\begin{figure}
\includegraphics[width=1\columnwidth]{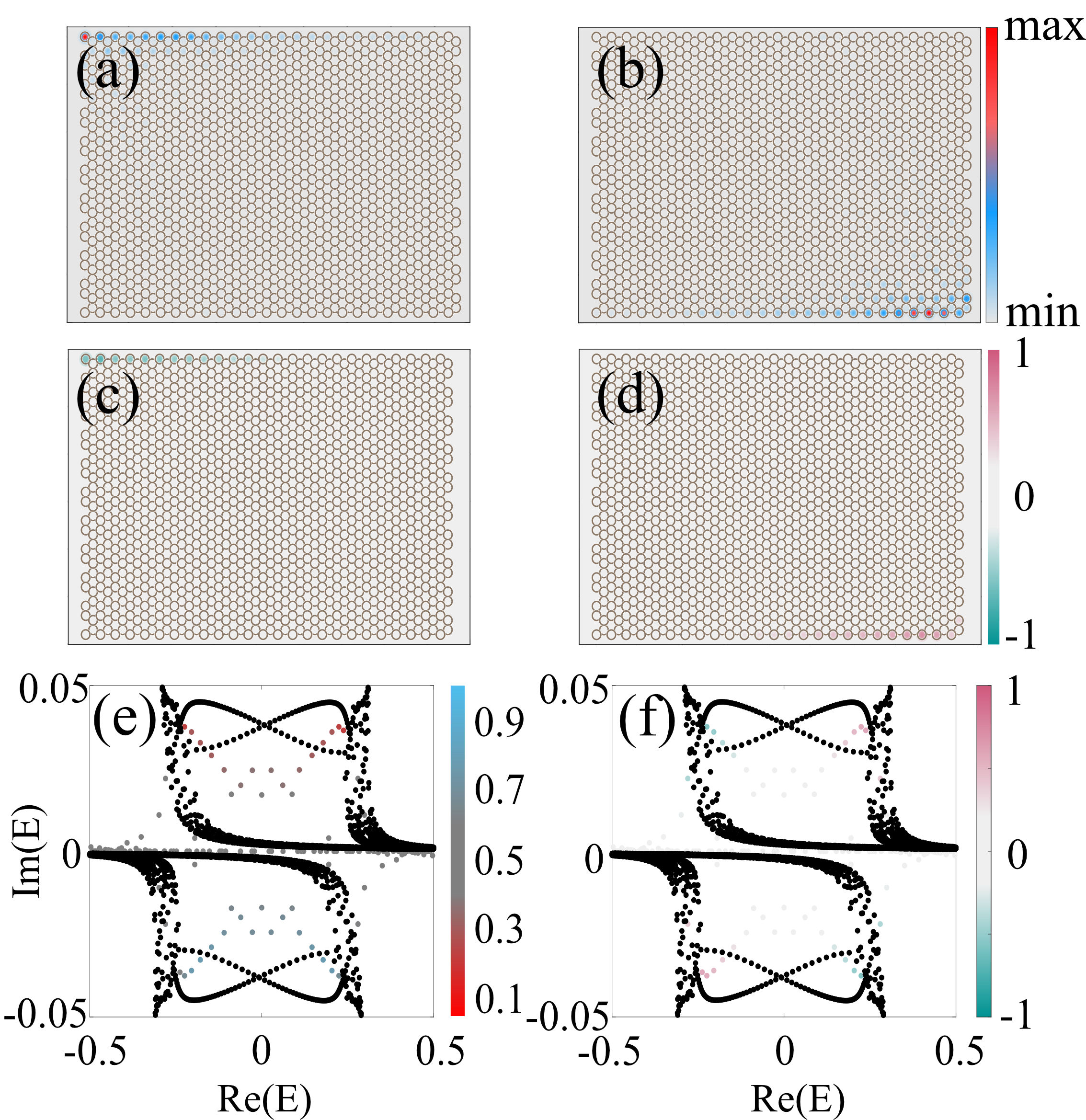}
\caption{ (a,b) Spatial distribution of HSAS, where they are now localized towards left (a) or right (b). (c-d) DOCP of HSAS; upper edge is spin "-" and lower edge is spin "+". (e) Energy spectrum of the considered HSAS system under both OBCs (color coded according to the expected position) and $x$-PBC/$y$-OBC (black points). (f) Energies from both OBCs are color coded according to the DOCP.  }
\label{Fig4}
\end{figure}
Having established the behaviour of the HSCS, we now turn to the antichiral edge state phase. In the Hermitian case, the edge states are localized at the upper and lower edges. When the sublattice-dependent non-Hermiticity is introduced, we find that the spatial distribution of edge states experiences drastic changes due to the NHSE and hybrid skin-antichiral edge states (HSAS) are formed. Now the states from the upper edge are localized towards the left (Fig.~\ref{Fig4}(a)), and states from lower edge are localized differently (Fig.~\ref{Fig4}(b)).

The topological origin of this phenomenon is suggested by the energy spectrum calculated from both boundary conditions (Fig.~\ref{Fig4}(e,f)) where topological states' energies from both OBCs are surrounded by $x$-PBC/$y$-OBC. It is further confirmed by the calculation of the winding number, which is -2 for the upper edge and 2 for the lower edge (see detailed calculation in~\cite{SM} ). These non-trivial topological invariants provide proof of the NHSE and explain why the two edge channels localize towards different directions of the system. This is consistent with the $\langle x \rangle$ results shown in Fig.~\ref{Fig4}(e). The edge states with $Im(E)>0$ are localized at top-left corner (red color coded); while the edge states with $Im(E)<0$ are localized at the bottom-right corner (blue color coded).

Furthermore, the localized HSAS inherit the spin-polarized  properties from the parent Hermitian edge states. By calculating the DOCP, we reveal the intrinsic spin properties: as shown in Fig. 4(c-d), states localized at the top-left corner are spin-down polarized; while the state at the other corner is spin-up polarized. This is also consistent with the previous band structure calculations. The emergent spin separation HSAS provides a link between non-Hermitian physics and spintronic polaritonics. The spatial distribution and spin properties in a corresponding continuous model agrees well with the tight binding results (see Supplementary Material~\cite{SM}).

Finally, it is essential to distinguish the corner-localized HSAS from other corner states. They are different from non-Hermitian corner states arising from higher-order skin effects, where all bulk states localize at corners and scale with the system size $\mathcal{O}$ $(N_x N_y)$~\cite{PhysRevB.106.L201302,Zhang:2021aa}. They are also distinct from Hermitian higher-order topological corner states, whose number is independent of system size ($\mathcal{O}$ $(1)$)~\cite{Imhof:2018aa,Serra-Garcia:2018aa,Peterson:2018aa,PhysRevLett.120.026801,Xue:2019aa,PhysRevLett.122.233902,PhysRevLett.124.063901,doi:10.1126/sciadv.adg4322}. In contrast, the number of HSAS scales with $\mathcal{O}$ $(N_x)$.

\emph{\textcolor{red}{Discussion.}}
In this work, we have theoretically proposed and analyzed the emergence of HSTS in exciton polariton honeycomb lattices induced by sublattice-dependent non-Hermiticity. By coupling the NHSE with the system’s intrinsic topological edge states, we demonstrate that gain–loss imbalance can relocalize topological states without destroying their spin polarization.

In the Chern insulator regime, the HSCS exhibit switchable localization that can be reversed by tuning the TE–TM splitting. This controllable transition, characterized by a change in the spectral winding number from $w=2$ to $w=-1$, provides a robust mechanism for manipulating non-reciprocal transport in driven dissipative systems.

Next, we showed the spin-polarized HSAS, which provides a link between the NHSE and spin degree of freedom. The resulting spin-up and spin-down polarized states are spatially separated, which may find application in spin-polarized lasing, non-reciprocal spin transport and measurement of spin-dependent polariton interactions. Importantly, the hybrid edge states remain robust against disorder and boundary defects (see Supplementary Material~\cite{SM}).

The agreement between tight binding and continuous model calculations supports the feasibility of our proposal using current polariton microcavity technologies (see Supplementary Material~\cite{SM}).
These results could motivate future studies on the interplay between HSTS and non-Hermitian nonlinear phenomena, paving the way to active control of non-Hermitian topological photonic and polaritonic devices.

\emph{\textcolor{red}{Acknowledgement.}}
The work was supported by the National Key R\&D Program of China (Grants No. 2024YFA1409600 and No. 2024YFE0213500), the National Natural Science Foundation of China (Grants No. 12374253, No. 12374174, and No. 12004064), R\&D project of Joint Funds of Liaoning Province (2023JH2/101800038), the Ministry of Education, Singapore (Grant No. MOE-MOET32023-0003).

\bibliography{main}

\end{document}